\documentclass{article}
\usepackage{spconf,amsmath,graphicx,hyperref}
\usepackage{array}
\usepackage{booktabs}
\usepackage{amssymb} 
\usepackage{subcaption} 
\usepackage{multirow}
\usepackage{ragged2e}   

\title{SynParaSpeech: Automated Synthesis of Paralinguistic Datasets for Speech Generation and Understanding}
\name{%
  \parbox{\linewidth}{%
    \centering
    \itshape
    Bingsong Bai$^{1\star}$ \qquad 
    Qihang Lu$^{1\star}$ \qquad
    Wenbing Yang$^{1\star}$ \qquad
    Zihan Sun$^{2}$ \qquad \\[2pt]
    Yueran Hou$^{2}$ \qquad
    Peilei Jia$^{2}$ \qquad 
    Songbai Pu$^{2}$ \qquad
    Ruibo Fu$^{3}$ \qquad \\[2pt]
    Yingming Gao$^{1}$ \qquad 
    Ya Li$^{1\dagger}$ \qquad
    Jun Gao$^{2\dagger}$
  }%
}
\address{
  \vspace{-1em}
  \thanks{$^{\star}$Equal contribution.       $^{\dagger}$Corresponding authors.} \\ 
  $^{1}$School of Artificial Intelligence, Beijing University of Posts and Telecommunications, China \\
  $^{2}$Hello Group Inc., China $^3$Institute of Automation, Chinese
 Academy of Sciences, China
}
\begin{document}
\ninept
\maketitle
\begin{abstract}
Paralinguistic sounds, like laughter and sighs, are crucial for synthesizing more realistic and engaging speech. However, existing methods typically depend on proprietary datasets, while publicly available resources often suffer from incomplete speech, inaccurate or missing timestamps, and limited real-world relevance. To address these problems, we propose an automated framework for generating large-scale paralinguistic data and apply it to construct the SynParaSpeech dataset. The dataset comprises 6 paralinguistic categories with 118.75 hours of data and precise timestamps, all derived from natural conversational speech. Our contributions lie in introducing the first automated method for constructing large-scale paralinguistic datasets and releasing the SynParaSpeech corpus, which advances speech generation through more natural paralinguistic synthesis and enhances speech understanding by improving paralinguistic event detection. The dataset and audio samples are available at \url{https://github.com/ShawnPi233/SynParaSpeech}.
\end{abstract}
\section{Introduction}
\label{sec:intro}

With rapid advances in deep learning, fields such as text-to-speech (TTS) and speech-language models (SLMs) have achieved high-quality speech synthesis. Most previous methods focus on semantic content and often ignore paralinguistic sounds. Yet in natural conversation, sounds like laughter and sighs are common. Recently, more researchers have aimed to improve paralinguistic speech synthesis to enhance interactivity and realism \cite{du2024cosyvoice, li2024spontaneous, wu2025anchored, orpheus-tts2025}.

Although the mentioned methods can synthesize paralinguistic speech, they rely on proprietary datasets with paralinguistic annotations, which are not publicly available, limiting scalable research. Thus, open paralinguistic datasets are essential. As shown in Table \ref{tab:dataset}, current open source datasets can be divided into two types. The first type includes audio-only datasets of paralinguistic events, such as AudioSet \cite{gemmeke2017audio}, ESC-50 \cite{piczak2015esc}, VocalSound \cite{gong2022vocalsound}, and Nonspeech7k \cite{rashid2023nonspeech7k}. These datasets, designed for sound event recognition, cover various non-semantic events but lack speech, text, and precise timestamps, making them unsuitable for paralinguistic speech synthesis or event localization. The second type is paralinguistic speech datasets with both speech and text, including Switchboard \cite{godfrey1992switchboard}, Fisher Speech \cite{cieri2004fisher}, MagicData-RAMC \cite{yang2022magicdata-ramc}, NVS\cite{ye2025nv38k}, and NVSpeech \cite{liao2025nvspeech}. While these provide transcriptions and paralinguistic labels, they have some limitations. Switchboard and Fisher Speech have low sampling rates, Fisher Speech and MagicData-RAMC cover few event categories, and Switchboard and NVSpeech lack precise timestamps. NVS consists mainly of animations, films, and shows whose non-verbal expressions are exaggerated and over-represented, so it departs from naturally occurring conversation. Similarly, both NVS and NVSpeech construct paralinguistic datasets by training ASR models with manually annotated paralinguistic events and subsequently employing these models to label new recordings. However, such ASR-based approaches suffer from two major limitations: the imbalance of paralinguistic categories, which introduces bias into the training data, and the scarcity and high cost of paralinguistic speech data, which restrict scalability.

To address these challenges, we propose an automated approach for synthesizing large-scale paralinguistic speech data and construct the SynParaSpeech dataset. Unlike previous ASR-based expansion methods \cite{ye2025nv38k, liao2025nvspeech}, our approach uses paralinguistic audio events and non-performative speech recordings to automatically generate a large dataset with precise timestamp annotations. This method can be easily extended to various paralinguistic categories and languages, offering insights for building large-scale multilingual paralinguistic datasets. Our contributions are as follows:
\begin{enumerate}
\item To the best of our knowledge, we propose the first automated approach for synthesizing large-scale paralinguistic speech datasets.

\item We introduce SynParaSpeech, a Chinese speech dataset containing 6 paralinguistic categories, with precise timestamped transcriptions and a total duration of 118.75 hours.

\item We demonstrate the effectiveness of SynParaSpeech by fine-tuning CosyVoice2 and F5-TTS, achieving substantial improvements in paralinguistic speech generation.

\item We apply prompt tuning to models such as Qwen 2.5 Omni and Kimi Audio, enhancing the detection of paralinguistic events, and explore the impact of varying prompt context lengths.
\end{enumerate}
\vspace{-0.5em}
\begin{table*}[t]
\centering
\caption{Paralinguistic Datasets Comparison.}
\label{tab:dataset}
\begin{tabular*}{\textwidth}{@{\extracolsep{\fill}}lcccccccc@{}}
\toprule
\textbf{Dataset} & \textbf{Hours} & \textbf{Clips} & \textbf{Type} & \textbf{Lang.} & \textbf{SR (kHz)} & \textbf{Timestamps} & \textbf{Speech} & \textbf{Available} \\ 
\midrule
AudioSet \cite{gemmeke2017audio} & 72.3 & 26,088 & 18 & - & - & \texttimes & \texttimes & \checkmark \\
ESC-50 \cite{piczak2015esc} & 0.33 & 240 & 10 & - & 22.05 & \texttimes & \texttimes & \checkmark \\
VocalSound \cite{gong2022vocalsound} & 20.46 & 21,024 & 6 & - & 16/44.1 & \texttimes & \texttimes & \checkmark \\
Nonspeech7k \cite{rashid2023nonspeech7k} & 6.75 & 7,014 & 7 & - & 32 & \texttimes & \texttimes & \checkmark \\
\midrule
Switchboard \cite{godfrey1992switchboard} & 260 & 11,699 & 42 & En & 8 & \texttimes & \checkmark & \checkmark \\
Fisher Speech \cite{cieri2004fisher} & 984 & 5,850 & 2 & En & 8 & \checkmark & \checkmark & \checkmark \\
MagicData-RAMC \cite{yang2022magicdata-ramc} & 180 & 219,325 & 3 & Zh & 16 & \checkmark & \checkmark & \checkmark \\
NVS\cite{ye2025nv38k} & 131 & 38,718 & 10 & Zh/En & 24 & \checkmark & \checkmark & \checkmark \\
NVSpeech \cite{liao2025nvspeech} &  573.4 & 174,179 & 18 & Zh & - & \texttimes & \checkmark & \checkmark \\
\midrule
SynParaSpeech (Ours) & 118.75 & 79,986 & 6 & Zh & 24 & \checkmark & \checkmark & \checkmark \\
\bottomrule
\end{tabular*}
\end{table*}
\vspace{-0.5mm}
\section{Methods}
\label{sec:methods}

\begin{figure*}[t] 
    \centering
    \includegraphics[width=\textwidth]{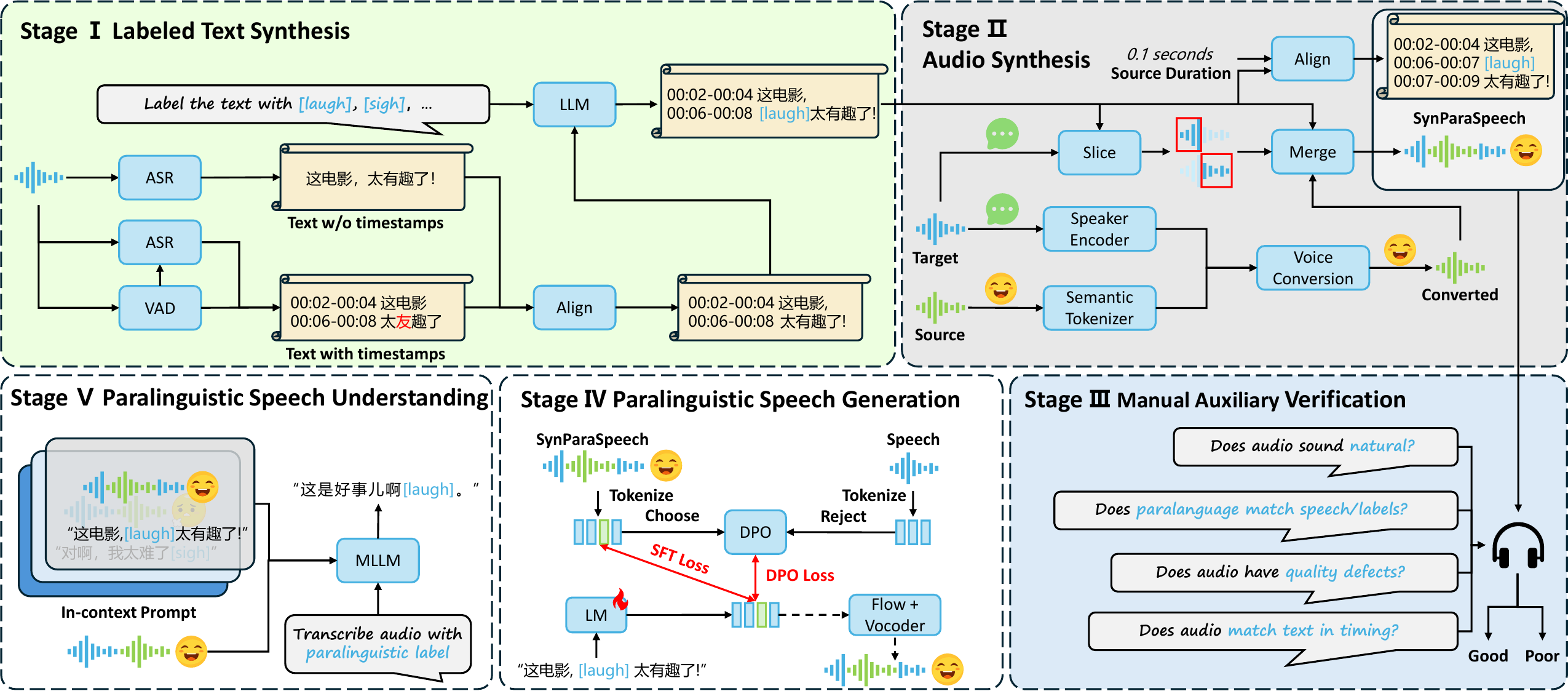}
    \caption{Overview of SynParaSpeech. (1) Labeled text with paralinguistic timestamps is synthesized. (2) Audio is synthesized and aligned with paralinguistic information. (3) The synthesized audio is verified across multiple dimensions. (4) Paralinguistic speech is generated via optimized modeling. (5) Paralinguistic speech understanding is achieved with in-context prompting.}
    \label{fig:synparaspeech}
\end{figure*}
    
Below, we describe how we built our dataset and the paralinguistic speech generation and understanding systems based on it. The construction of the dataset includes creating paralinguistic labeled text, synthesizing paralinguistic speech audio, and manual checks, as shown in Fig. \ref{fig:synparaspeech}.

\subsection{SynParaSpeech Labeled Text Synthesis}
First we designed a processing pipeline to generate accurate transcriptions aligned timestamps and paralinguistic tags. The audio was processed in two parallel steps: (1) three ASR models Whisper Large V3 \cite{radford2023whisper}, SenseVoice \cite{an2024funaudiollm} and Paraformer \cite{gao2022paraformer} were applied to obtain sentence-level transcriptions through majority voting. (2) Voice Activity Detection (VAD) and segmented the audio into shorter clips.

Since ASR performs poorly on very short clips, we validated each VAD-determined split point by splitting the audio at these points into left and right sub-clips, transcribing each sub-clip using the same ASR models, and calculating the edit distance between these sub-clip transcriptions and the full-sentence transcription to determine how the sub-clip texts aligned within the complete sentence. For each split point, we took the transcription of the longer sub-clip, computed its Character Error Rate (CER) against the corresponding segment of the full-sentence transcription, and deemed the split accurate (yielding timestamped text clips) if the CER was below 0.1. Final alignment then leveraged these VAD outputs and ASR results, with verification via Stable Whisper \cite{stable-ts2023}, to generate precisely timestamped text segments.

To incorporate paralinguistic tags, we used a mainstream large language model (LLM) Deepseek V3 \cite{liu2024deepseek} to build a tagged text dataset. We input the transcriptions into the LLM, prompting it to select the most appropriate tag from \(\texttt{[laugh]}\), \(\texttt{[sigh]}\), \(\texttt{[gasp]}\), \(\texttt{[throat clearing]}\), \(\texttt{[pause]}\), \(\texttt{[tsk]}\) without altering the original text and insert it at the boundary of relevant text segments. Ultimately, this process yielded a text dataset annotated with paralinguistic tags.

\subsection{SynParaSpeech Audio Synthesis}
\label{sec:concat_pipeline}
The objective of the audio synthesis phase (Stage II) is to generate speech containing paralinguistic cues that conforms to the paralinguistic-annotated text produced in Stage I, by integrating established synthesis methodologies for paralinguistic audio and speech audio.  

First, using the paralinguistic label from Stage I (e.g., \texttt{[laugh]} as in Fig. \ref{fig:synparaspeech}), we randomly pick an audio clip from the matching paralinguistic audio corpus. To keep timbre consistent between paralinguistic audio and speech, we conduct voice conversion (VC): paralinguistic audio is the source, speech audio the target. We use ASR model Whisper Large V3 \cite{radford2023whisper} to encode the source audio's semantic content, and speaker encoder CAM++ \cite{wang2023cam++} to extract the target's speaker embedding. These features go into zero-shot VC model SeedVC \cite{liu2024seedvc} to get paralinguistic audio with adjusted timbre. 

Additionally, speech audio is sliced based on the timestamped text from Stage I. Then, the timbre-converted paralinguistic audio is inserted into the corresponding positions of the speech segments. Finally, all processed speech segments are merged in chronological order to get the final speech.
\subsection{Manual Auxiliary Verification}
\label{sec:human_annoate}
Although the proposed process enables efficient automated synthesis of paralinguistic speech, additional manual verification was conducted to ensure alignment with human perception in terms of naturalness, audio quality, and paralinguistic categories. Professionals in speech were invited to evaluate and refine the synthesized SynParaSpeech. The evaluation considered four aspects: the naturalness of the audio, with emphasis on fluent transitions between speech and paralinguistic cues; the quality of paralinguistic expressions, including timbre consistency and accurate label matching; the overall audio quality, ensuring the absence of noise, clipping, or distortion; and the accuracy of temporal alignment between audio and text, avoiding missing, incorrect, or redundant characters. After refinement, only compliant audio clips were retained.

\section{SynParaSpeech Dataset}
\label{sec:statics}

\begin{table}[!t]
\centering
\setlength\tabcolsep{5pt}   
\caption{Statistics of the SynParaSpeech Dataset.}
\label{tab:statics}
\begin{tabular}{lcccc}
\toprule
\textbf{Category} & \textbf{Hours} & \textbf{Clips} & \textbf{Avg.(s)} & \textbf{Share}\\
\midrule
Sigh            & 28.22 & 17,706 & 5.74 & 23.76 \% \\
Throat Clearing & 25.45 & 18,827 & 4.87 & 21.43 \% \\
Laugh           & 20.84 & 13,023 & 5.76 & 17.55 \% \\
Pause           & 18.30 &  9,643 & 6.83 & 15.41 \% \\
Tsk             & 14.82 & 11,941 & 4.47 & 12.48 \% \\
Gasp            & 11.11 &  8,846 & 4.52 &  9.36 \% \\
\midrule
Total & 118.75 & 79,986 & 5.34 & 100.00 \%\\
\bottomrule
\end{tabular}
\end{table}

\begin{table*}[ht]
\centering
\caption{Paralinguistic TTS with SynParaSpeech dataset. Confidence interval for MOS scores is 95\%.}
\label{tab:tts_experiment}
\begin{tabular*}{\textwidth}{@{\extracolsep{\fill}}l|cccc|cccc}
\toprule
\textbf{Model}               & \textbf{PMOS} $\uparrow$ & \textbf{NMOS} $\uparrow$ & \textbf{SMOS} $\uparrow$ & \textbf{QMOS} $\uparrow$ & \textbf{CER(\%)} $\downarrow$ & \textbf{SECS} $\uparrow$ & \textbf{UTMOSv2} $\uparrow$ \\
\midrule
F5-TTS (Baseline)            & 1.16 ± 0.01 & 4.08 ± 0.02 & \textbf{4.52 ± 0.02} & 3.95 ± 0.03 & \textbf{6.01} & \textbf{0.76} & 3.01    \\
\quad+ NVS SFT               & 1.49 ± 0.03 & 3.83 ± 0.03 & 4.03 ± 0.02 & 3.75 ± 0.03 & 12.56 & \underline{0.74} & 3.01    \\
\quad+ SynParaSpeech SFT     & 3.10 ± 0.04 & 4.16 ± 0.02 & \underline{4.41 ± 0.02} & 4.08 ± 0.02 & 7.26 & \underline{0.74} & 2.83    \\
\midrule
CosyVoice2 (Baseline)        & 1.88 ± 0.04 & \textbf{4.24 ± 0.02} & 3.71 ± 0.03 & 4.00 ± 0.03 & \underline{6.58} & 0.70 & \textbf{3.13}    \\
\quad+ NVS SFT               & 2.35 ± 0.05 & 4.06 ± 0.02 & 3.47 ± 0.03 & 3.95 ± 0.03 & 9.50 & 0.69 & \underline{3.02}    \\
\quad+ SynParaSpeech SFT     & 3.31 ± 0.04 & 4.11 ± 0.02 & 3.74 ± 0.03 & 4.01 ± 0.02 & 11.00 & 0.71 & 2.78    \\
\qquad+ DPO-Staged         & \underline{3.40 ± 0.04} & 4.15 ± 0.02 & 3.84 ± 0.02 & \underline{4.09 ± 0.02} & 10.91 & 0.70 & 2.87    \\
\qquad+ DPO-Joint          & \textbf{3.46 ± 0.04} & \underline{4.17 ± 0.02} & 4.03 ± 0.03 & \textbf{4.12 ± 0.02} & 11.78 & 0.71 & 2.83    \\
\bottomrule
\end{tabular*}
\end{table*}

As shown in Table \ref{tab:statics}, the SynParaSpeech dataset includes speech data for six paralinguistic events: sigh, throat clearing, laugh, pause, tsk, and gasp. Its statistical metrics cover total duration (Hours), number of clips (Clips), average clip duration (Avg.(s)), and proportion in the full dataset (Share) for each event.  

Since the natural distribution of semantic scenarios affects paralinguistic event frequency, we did not enforce equal proportions for each category during dataset construction. Even so, statistics still show good balance across all categories: the highest proportion (sigh, 23.76\%) and lowest proportion (gasp, 9.36\%) differ within a reasonable range, and no category is overly dominant or extremely scarce (sample sizes range from 8,846 to 18,827). Thus, the dataset can provide balanced multi-category support for paralinguistic event analysis and modeling.  

\section{Experiments}
\label{sec:experiments}
\subsection{Experimental Setups}
\label{sec:experiment_setup}

To evaluate the effectiveness of SynParaSpeech in both paralinguistic speech synthesis and speech understanding tasks, we conducted experiments on paralinguistic TTS and event detection.

\subsubsection{Paralinguistic TTS}
\label{sec:experiment_tts}

In the paralinguistic TTS experiments, we employed SynParaSpeech to conduct supervised fine-tuning (SFT) on state-of-the-art open-source TTS models, namely the autoregressive CosyVoice2 \cite{du2024cosyvoice} and the non-autoregressive F5-TTS \cite{chen2024f5}. For comparison, we also considered the NVS dataset \cite{ye2025nv38k}, a large and diverse paralinguistic resource collected from real-world media and automatically annotated. To ensure comparability, we selected four common paralinguistic labels—\texttt{[laugh]}, \texttt{[sigh]}, \texttt{[gasp]}, and \texttt{[throat clearing]}—that are supported by both SynParaSpeech and NVS. We kept 2\% of the data randomly as a validation set, the rest for training. For evaluation, the LLM DeepSeek V3 \cite{liu2024deepseek} was used to rewrite Seed-TTS-Eval \cite{anastassiou2024seed} texts (with paralinguistic labels inserted), making 100 inference texts per category.

We compared the CosyVoice2 and F5-TTS checkpoints with their fine-tuned derivatives trained on SynParaSpeech and on NVS. To overcome the issue that SFT only focuses on the model’s overall output loss and lacks focus on paralinguistic synthesis, we applied direct preference optimization (DPO) \cite{rafailov2023dpo} to CosyVoice2 in two configurations: DPO-Staged (DPO after SFT) and DPO-Joint (optimizing both objectives together). Leveraging the clean origin of SynParaSpeech, each utterance naturally provides paired positives (with paralinguistic events) and negatives (without). During DPO training, the original utterance serves as the rejected sample $\mu_{\mathrm{o}}$, while its SynParaSpeech counterpart is the chosen sample $\mu_{\mathrm{p}}$. The preference learning objective is thus defined as

\begin{equation}
\mathcal{L}_{\text{dpo}} = -\mathbb{E}_y \left[ \log \sigma \left( \beta \cdot \log \frac{\pi_\theta(\mu_p \mid y) \cdot \pi_{\text{ref}}(\mu_o \mid y)}{\pi_\theta(\mu_o \mid y) \cdot \pi_{\text{ref}}(\mu_p \mid y)} \right) \right]
\end{equation}
where $\pi_{\theta}$ is the target model, $\pi_{\mathrm{ref}}$ the reference model, $\beta=0.01$ the temperature, and $\sigma(\cdot)$ the sigmoid function.

For training, CosyVoice2’s language model is optimized with Adam \cite{kinga2015adam} at a learning rate of $1\times10^{-5}$ for a maximum of 50 epochs, with early stopping (patience 10), a constant learning rate, and a 2500-step linear warm-up. Training stability is maintained via gradient clipping (threshold 5), gradient accumulation over 2 steps, and dynamic batching with a maximum of 2000 frames. Following the NVS setup, F5-TTS is trained for 400 epochs with a learning rate of $1\times10^{-4}$, a cosine annealing scheduler with 1000 warm-up updates, and a frame-based batch size of 30,000 per GPU. To initialize new paralinguistic tokens, we use embeddings from the RoBERTa-based \cite{liu2019roberta} text encoder of CLAP \cite{elizalde2023clap}, aligning dimensions via interpolation.

Model performance was assessed using both objective and subjective metrics. Objective measures include character error rate (CER) for intelligibility, SECS for speaker similarity, and UTMOSv2 \cite{baba2024utmosv2} for speech quality. Subjective evaluation employed 5-point ratings: PMOS for the paralinguistic quality, NMOS for naturalness, SMOS for speaker similarity, and QMOS for overall audio quality. For MOS ratings, 21 volunteers participated in a double-blind evaluation.

\vspace{-1em}
\subsubsection{Paralinguistic Event Detection}
\label{sec:experiment_ved}
\begin{table}[!t]
\centering
\vspace{-1em}
\caption{Paralinguistic event detection results.}
\label{tab:vocal_event_experiment}
\begin{tabular}{>{\RaggedRight}p{0.7cm}c|ccc}
\toprule
\textbf{Model} & \textbf{Context} & \textbf{Acc.\ $\uparrow$} & \textbf{F1 Score\ $\uparrow$} & \textbf{CER\,(\%)\ $\downarrow$} \\
\midrule
\multirow{5}{=}{Kimi Audio} 
& $-$   & 0.320 & 0.294 & 17.79 \\ 
& 1     & 0.314 & 0.312 & 11.30 \\ 
& 3     & 0.354 & 0.336 & \textbf{10.61} \\ 
& 5     & \textbf{0.382} & \textbf{0.340} & 11.11 \\ 
& 7     & 0.371 & 0.331 & 11.01 \\
\midrule
\multirow{5}{=}{Qwen 2.5 Omni} 
& $-$   & 0.215 & 0.189 & 23.52 \\ 
& 1     & 0.337 & 0.357 & 21.18 \\ 
& 3     & 0.460 & 0.447 & 20.60 \\ 
& 5     & \textbf{0.473} & \textbf{0.471} & \textbf{19.48} \\ 
& 7     & 0.423 & 0.362 & 20.07 \\ 
\bottomrule
\end{tabular}
\vspace{-1em}
\end{table}
In the paralinguistic event detection experiments, we applied SynParaSpeech-based prompt tuning to multimodal large language models (MLLMs) and analyzed the effect of context size. Following MMSU \cite{wang2025mmsu}, Kimi Audio \cite{ding2025kimi} was chosen for its strong paralinguistic reasoning, and Qwen 2.5 Omni \cite{xu2025qwen2} for its superior paralinguistic perception. Both models were evaluated with and without SynParaSpeech prompt tuning. Each paralinguistic category used 100 test pairs with varying context prompts. Performance was measured using accuracy, macro F1 score, and CER.

\subsection{Experimental Results and Analysis}
\label{sec:experiment_result}
\subsubsection{Paralinguistic TTS  Experimental Results}
\label{sec:result_tts}
The experimental results of paralinguistic TTS are shown in Table \ref{tab:tts_experiment}. Comparing F5-TTS across the two datasets, the model fine-tuned on SynParaSpeech shows improvements in paralinguistic synthesis, naturalness, and overall quality. The gain is most evident in paralinguistic synthesis, where PMOS increases significantly compared with the model fine-tuned on NVS. SynParaSpeech also provides advantages in naturalness, speaker similarity, overall quality, and articulation. Similarly, CosyVoice2 fine-tuned on SynParaSpeech achieves a large improvement in paralinguistic synthesis, a moderate gain in speaker similarity, and overall advantages over the NVS-fine-tuned counterpart in paralinguistic quality, naturalness, speaker similarity, and overall quality. These results confirm that SynParaSpeech can enhance paralinguistic synthesis while preserving naturalness, speaker similarity, and overall speech quality.

Furthermore, F5-TTS uses non-autoregressive flow matching, it achieves high naturalness and speaker similarity. In contrast, CosyVoice2, with its autoregressive multi-stage design, better models paralinguistic features and produces higher audio quality. Building on this, we investigated how DPO enables CosyVoice2 to better capture paralinguistic details and further enhance its paralinguistic capability. The last three rows show that DPO boosts paralinguistic quality, naturalness, and speaker similarity. For both DPO types, our joint SFT training beats staged training, increases all subjective scores, and achieves the best PMOS and QMOS.

We also observed that models trained on paralinguistic datasets show slight declines in objective metrics CER, SECS, and UTMOSv2, as these metrics are designed for standard speech. For example, laughter may be transcribed as “ha ha”, raising CER. Nevertheless, models trained on SynParaSpeech still maintain high NMOS and QMOS while achieving clear PMOS gains.
\vspace{-1em}
\subsubsection{Paralinguistic Event Detection Experimental Results}
Table \ref{tab:vocal_event_experiment} shows the impact of SynParaSpeech prompt tuning and context quantity on paralinguistic event detection. Notably, due to Kimi Audio’s poor support for long-context input versus Qwen 2.5 Omni’s strong support, the context quantity for Kimi Audio refers to the total number of prompts, while that for Qwen 2.5 Omni corresponds to the number per paralinguistic category. SynParaSpeech consistently improves accuracy and macro F1 over the no-context baseline for both models, confirming its value for paralinguistic perception and reasoning. Context quantity also matters: performance improves with more examples, peaking at 5-shot; beyond this (e.g., 7-shot), the improvements stop or even get worse. This reflects a trade-off between informative prompts and avoiding input overload. Overall, SynParaSpeech enhances paralinguistic event detection capability, and an optimal context size is essential to maximize its benefits.
\section{Conclusion}
\label{sec:conclusion}
This paper proposes an automated approach for large-scale paralinguistic dataset synthesis and introduces the SynParaSpeech dataset. SynParaSpeech, featuring multiple paralinguistic categories with fine-grained annotations and a substantial scale, serves as an effective resource to facilitate both paralinguistic speech synthesis (TTS) and paralinguistic event detection tasks. Experimental results demonstrate that integrating SynParaSpeech enhances the generation quality of TTS models and boosts the performance of paralinguistic event detection models.

\section{Acknowledgement}
\label{sec:acknowledgement}
The work was supported by the National Natural Science Foundation of China (NSFC) (No. 62271083),  the Key Project of the National Language Commission (No. ZDI145-81), and the Fundamental Research Funds for the Central Universities (No. 2023RC73).


\clearpage

\bibliographystyle{IEEEbib}
\bibliography{strings}

\begin{thebibliography}{10}

\bibitem{du2024cosyvoice}
Zhihao Du, Yuxuan Wang, Qian Chen, Xian Shi, Xiang Lv, et~al.,
\newblock ``Cosyvoice 2: Scalable streaming speech synthesis with large language models,''
\newblock {\em arXiv preprint arXiv:2412.10117}, 2024.

\bibitem{li2024spontaneous}
Weiqin Li, Peiji Yang, Yicheng Zhong, Yixuan Zhou, Zhisheng Wang, Zhiyong Wu, Xixin Wu, and Helen Meng,
\newblock ``Spontaneous style text-to-speech synthesis with controllable spontaneous behaviors based on language models,''
\newblock in {\em Proc. Interspeech 2024}, 2024, pp. 1785--1789.

\bibitem{wu2025anchored}
Ning-Qian Wu, Ya-Jun Hu, Liping Chen, and Zhen-Hua Ling,
\newblock ``Anchored monotonic alignment and representation substitution for rare spontaneous behaviors in spontaneous speech synthesis,''
\newblock in {\em ICASSP 2025}. IEEE, 2025, pp. 1--5.

\bibitem{orpheus-tts2025}
canopyai,
\newblock ``Orpheus tts,'' \url{https://github.com/canopyai/Orpheus-TTS}, 2025.

\bibitem{gemmeke2017audio}
Jort~F Gemmeke, Daniel~PW Ellis, Dylan Freedman, Aren Jansen, Wade Lawrence, R~Channing Moore, Manoj Plakal, and Marvin Ritter,
\newblock ``Audio set: An ontology and human-labeled dataset for audio events,''
\newblock in {\em ICASSP 2017}. IEEE, 2017, pp. 776--780.

\bibitem{piczak2015esc}
Karol~J Piczak,
\newblock ``Esc: Dataset for environmental sound classification,''
\newblock in {\em Proceedings of the 23rd ACM international conference on Multimedia}, 2015, pp. 1015--1018.

\bibitem{gong2022vocalsound}
Yuan Gong, Jin Yu, and James Glass,
\newblock ``Vocalsound: A dataset for improving human vocal sounds recognition,''
\newblock in {\em ICASSP 2022}. IEEE, 2022, pp. 151--155.

\bibitem{rashid2023nonspeech7k}
Muhammad~Mamunur Rashid, Guiqing Li, and Chengrui Du,
\newblock ``Nonspeech7k dataset: Classification and analysis of human non-speech sound,''
\newblock {\em IET Signal Processing}, vol. 17, no. 6, pp. e12233, 2023.

\bibitem{godfrey1992switchboard}
John~J Godfrey, Edward~C Holliman, and Jane McDaniel,
\newblock ``Switchboard: Telephone speech corpus for research and development,''
\newblock in {\em Acoustics, speech, and signal processing, ieee international conference on}. IEEE Computer Society, 1992, vol.~1, pp. 517--520.

\bibitem{cieri2004fisher}
Christopher Cieri, David Graff, Owen Kimball, Dave Miller, and Kevin Walker,
\newblock {\em Fisher English training speech part 1 transcripts},
\newblock Lead Discovery Center LDC, 2004.

\bibitem{yang2022magicdata-ramc}
Zehui Yang, Yifan Chen, Lei Luo, Runyan Yang, Lingxuan Ye, Gaofeng Cheng, Ji~Xu, Yaohui Jin, Qingqing Zhang, Pengyuan Zhang, et~al.,
\newblock ``Open source magicdata-ramc: A rich annotated mandarin conversational (ramc) speech dataset,''
\newblock in {\em Proc. Interspeech 2022}, 2022, pp. 1736--1740.

\bibitem{ye2025nv38k}
Runchuan Ye, Yixuan Zhou, Renjie Yu, Zijian Lin, Kehan Li, Xiang Li, Xin Liu, Guoyang Zeng, and Zhiyong Wu,
\newblock ``A scalable pipeline for enabling non-verbal speech generation and understanding,''
\newblock {\em arXiv preprint arXiv:2508.05385}, 2025.

\bibitem{liao2025nvspeech}
Huan Liao, Qinke Ni, Yuancheng Wang, Yiheng Lu, Haoyue Zhan, Pengyuan Xie, Qiang Zhang, and Zhizheng Wu,
\newblock ``Nvspeech: An integrated and scalable pipeline for human-like speech modeling with paralinguistic vocalizations,''
\newblock {\em arXiv preprint arXiv:2508.04195}, 2025.

\bibitem{radford2023whisper}
Alec Radford, Jong~Wook Kim, Tao Xu, Greg Brockman, Christine McLeavey, and Ilya Sutskever,
\newblock ``Robust speech recognition via large-scale weak supervision,''
\newblock in {\em ICML}. PMLR, 2023, pp. 28492--28518.

\bibitem{an2024funaudiollm}
Keyu An, Qian Chen, Chong Deng, Zhihao Du, Changfeng Gao, Zhifu Gao, et~al.,
\newblock ``Funaudiollm: Voice understanding and generation foundation models for natural interaction between humans and llms,''
\newblock {\em arXiv preprint arXiv:2407.04051}, 2024.

\bibitem{gao2022paraformer}
Zhifu Gao, Shiliang Zhang, Ian McLoughlin, and Zhijie Yan,
\newblock ``Paraformer: Fast and accurate parallel transformer for non-autoregressive end-to-end speech recognition,''
\newblock in {\em Proc. Interspeech 2022}, 2022, pp. 2063--2067.

\bibitem{stable-ts2023}
jianfch,
\newblock ``Stabilizing timestamps for whisper,'' \url{https://github.com/jianfch/stable-ts}, 2023.

\bibitem{liu2024deepseek}
Aixin Liu, Bei Feng, Bing Xue, Bingxuan Wang, Bochao Wu, Chengda Lu, Chenggang Zhao, Chengqi Deng, Chenyu Zhang, Chong Ruan, et~al.,
\newblock ``Deepseek-v3 technical report,''
\newblock {\em arXiv preprint arXiv:2412.19437}, 2024.

\bibitem{wang2023cam++}
Hui Wang, Siqi Zheng, Yafeng Chen, Luyao Cheng, and Qian Chen,
\newblock ``Cam++: A fast and efficient network for speaker verification using context-aware masking,''
\newblock in {\em Proc. Interspeech 2023}, 2023, pp. 5301--5305.

\bibitem{liu2024seedvc}
Songting Liu,
\newblock ``Zero-shot voice conversion with diffusion transformers,''
\newblock {\em arXiv preprint arXiv:2411.09943}, 2024.

\bibitem{chen2024f5}
Yushen Chen, Zhikang Niu, Ziyang Ma, Keqi Deng, Chunhui Wang, Jian Zhao, Kai Yu, and Xie Chen,
\newblock ``F5-tts: A fairytaler that fakes fluent and faithful speech with flow matching,''
\newblock {\em arXiv preprint arXiv:2410.06885}, 2024.

\bibitem{anastassiou2024seed}
Philip Anastassiou, Jiawei Chen, Jitong Chen, et~al.,
\newblock ``Seed-tts: A family of high-quality versatile speech generation models,''
\newblock {\em arXiv preprint arXiv:2406.02430}, 2024.

\bibitem{rafailov2023dpo}
Rafael Rafailov, Archit Sharma, Eric Mitchell, Christopher~D Manning, Stefano Ermon, and Chelsea Finn,
\newblock ``Direct preference optimization: Your language model is secretly a reward model,''
\newblock {\em Advances in neural information processing systems}, vol. 36, pp. 53728--53741, 2023.

\bibitem{kinga2015adam}
Diederik Kinga, Jimmy~Ba Adam, et~al.,
\newblock ``A method for stochastic optimization,''
\newblock in {\em ICLR}. California;, 2015, vol.~5.

\bibitem{liu2019roberta}
Yinhan Liu, Myle Ott, Naman Goyal, Jingfei Du, Mandar Joshi, Danqi Chen, et~al.,
\newblock ``Roberta: A robustly optimized bert pretraining approach,''
\newblock {\em arXiv preprint arXiv:1907.11692}, 2019.

\bibitem{elizalde2023clap}
Benjamin Elizalde, Soham Deshmukh, Mahmoud Al~Ismail, and Huaming Wang,
\newblock ``Clap learning audio concepts from natural language supervision,''
\newblock in {\em ICASSP 2023}. IEEE, 2023, pp. 1--5.

\bibitem{baba2024utmosv2}
Kaito Baba, Wataru Nakata, Yuki Saito, and Hiroshi Saruwatari,
\newblock ``The t05 system for the {V}oice{MOS} {C}hallenge 2024: Transfer learning from deep image classifier to naturalness {MOS} prediction of high-quality synthetic speech,''
\newblock in {\em IEEE Spoken Language Technology Workshop (SLT)}, 2024.

\bibitem{wang2025mmsu}
Dingdong Wang, Jincenzi Wu, Junan Li, Dongchao Yang, Xueyuan Chen, Tianhua Zhang, and Helen Meng,
\newblock ``Mmsu: A massive multi-task spoken language understanding and reasoning benchmark,''
\newblock {\em arXiv preprint arXiv:2506.04779}, 2025.

\bibitem{ding2025kimi}
Ding Ding, Zeqian Ju, Yichong Leng, Songxiang Liu, Tong Liu, Zeyu Shang, Kai Shen, Wei Song, Xu~Tan, Heyi Tang, et~al.,
\newblock ``Kimi-audio technical report,''
\newblock {\em arXiv preprint arXiv:2504.18425}, 2025.

\bibitem{xu2025qwen2}
Jin Xu, Zhifang Guo, Jinzheng He, Hangrui Hu, Ting He, Shuai Bai, Keqin Chen, Jialin Wang, Yang Fan, Kai Dang, et~al.,
\newblock ``Qwen2. 5-omni technical report,''
\newblock {\em arXiv preprint arXiv:2503.20215}, 2025.

\end{thebibliography}

\end{document}